\documentstyle[10pt,aas2pp4,tighten]{article}

\received{}
\revised{}
\accepted{}
\journalid{}{}
\articleid{}{}
\paperid{}
\cpright{}{}
\ccc{}

\slugcomment{Submitted to the Astrophysical Journal}

\lefthead{Wood et al.}
\righthead{Diffuse H$\alpha$ Background}

\begin{document}

\title{A Model for the Scattered Light Contribution and Polarization of 
the Diffuse H$\alpha$ Galactic Background}

\author{Kenneth Wood\altaffilmark{1} and Ron Reynolds\altaffilmark{2}}

\altaffiltext{1}{Smithsonian Astrophysical Observatory,
60 Garden Street, Cambridge, MA 02138; kenny@claymore.harvard.edu}
\altaffiltext{2}{Astronomy Department, University of Wisconsin, 
475 North Charter Street, Madison, WI~53703; reynolds@astro.wisc.edu}

\authoremail{kenny@claymore.harvard.edu}

\begin{abstract}

We present Monte Carlo simulations of the Diffuse H$\alpha$ Galactic 
Background.  Our models comprise direct and multiply scattered H$\alpha$ 
radiation from the kpc scaleheight Warm Ionized Medium and midplane H~II 
regions.  The scattering is off dust that is assumed to be well mixed with 
the gas, with an axisymmetric density distribution taken from the literature.  
The results of our simulations are 
all-sky H$\alpha$ images that enable us to separate out the contributions 
of direct and scattered radiation.  We also determine how far the model 
H$\alpha$ photons have traveled, i.e., how far we see into the Galaxy 
at H$\alpha$.  Our models reproduce the overall characteristics of the 
observed H$\alpha$ background and predict the scattered H$\alpha$ intensity at 
high latitudes is in the range 5\% to 20\% of the total intensity, 
in agreement with estimations based on [S~II]/H$\alpha$ and 
[O~III]/H$\alpha$ line ratio 
measurements.  The polarization arising from dust scattering of H$\alpha$ 
from midplane H~II regions is predicted to be less than 1\% at high latitude.

\end{abstract}

\keywords{diffuse radiation --- H~II regions --- ISM --- 
radiation transfer --- dust scattering --- polarization}

\section{Introduction}

Ubiquitous but faint H$\alpha$ proves the presence of the Warm 
Ionized Medium (WIM), a major but poorly understood component of the 
ISM containing the majority of the ionized gas in the Galaxy (see a 
review by Reynolds et al. 1995).  
The ionized gas is seen at all galactic latitudes 
and has a scale height of about 900pc (judged from the dispersion measure 
of the radio signals from pulsars in globular clusters), a temperature of 
around 10000K, and volume filling factor in the range 0.1 to 0.4.  In the 
solar neighborhood, the total power represented by the 
ionizations producing the H$\alpha$ is about 15\% of the total 
ionizing rate from OB stars, or about 100\% of the power 
ejected by supernovae.  The source of ionization is not well understood and 
has been narrowed down to either (a) the leakage of the ionizing photons from 
OB stars to the diffuse ionized gas, which actually provides the correct
spectrum of the O, N, and S ions (Domg\"{o}rgen and Mathis 1994; see also
Dove \& Shull 1994; Dove, Shull, \& Ferrara 1999;  Miller \& Cox 1993; 
Elmegreen 1998); or (b) 
some unknown, more distributed source of ionization (Sciama 1995).  

Low spatial resolution mapping of the H$\alpha$ emission (Reynolds 1990) 
has revealed the following characteristics.  

\begin{itemize}

\item At low latitudes ($|b| < 10^\circ$) the Galactic midplane is dominated 
by H$\alpha$ emission from bright H~II regions.   

\item At intermediate Galactic latitudes ($10^\circ < |b| < 50^\circ$) the 
H$\alpha$ intensity follows a $\csc |b|$ law with the 
average intensity being $<I_\alpha\sin |b|> \approx 1R$, where 
$R = 10^6/4\pi$ H$\alpha$ photons s$^{-1}$ cm$^{-2}$ steradian$^{-1}$.

\item Above $|b|=50^\circ$ the emissivity appears to be about a factor of 
two below the $\csc |b|$ law for lower latitudes.

\item Localized departures from the $\csc |b|$ law are evident at all 
latitudes, indicating the clumpy and filamentary nature of the WIM.  These 
features are becoming more evident through early results from the 
Wisconsin H$\alpha$ Mapper (WHAM) all sky survey (Reynolds et al. 1998; 
Haffner, Reynolds, \& Tufte 1998).

\end{itemize}

The Galactic contributions to the H$\alpha$ background are direct and 
scattered H$\alpha$ from the WIM, direct H$\alpha$ from midplane H~II regions, 
and H$\alpha$ from H~II regions that is scattered by dust at high latitudes.  
Observations of [S~II]/H$\alpha$ line ratios along 
sightlines at low latitude (towards midplane H~II regions) and at high 
latitudes in the diffuse ISM (Reynolds 1988), show that the line ratios in 
the diffuse ISM are larger by up to a factor of ten than those in traditional 
H~II regions.  This indicates 
that the ionization state of the WIM is very 
different from that of H~II regions, and based on these observations, 
Reynolds (1990) concluded that the scattered 
component from midplane H~II regions was at most 15\% of the total H$\alpha$ 
intensity.  

Ferrara et al. (1996) presented a radiation transfer model for H$\alpha$ 
emission and scattering in the edge-on galaxy NGC~891.  They found that at 
600pc from the midplane, the scattered light from H~II regions is around 
10\% of the total H$\alpha$ intensity.  Their models also predict a 
dust scattered polarization of less than 1\%.  The analysis that we 
present in this paper is very similar to that of Ferrara et al. (1996) 
except we are now viewing the radiation transfer simulation from inside 
the Galaxy.  Our goal is to determine, 
from a theoretical/modeling perspective, the relative contributions of 
direct and scattered H$\alpha$ at high latitudes, 
how large a region of the ISM contributes 
to the observed intensity, and whether our models can reproduce the general 
characteristics of the H$\alpha$ background.  In addition, we present 
a prediction of the polarization arising from dust scattering of 
the H$\alpha$ emission.  The ingredients of our models are presented in 
\S 2, and the scattered light models and polarization predictions are 
presented in \S3 and \S4.

\section{Models}

We perform the radiation transfer with a 3D Monte Carlo scattering code.  
We have previously modeled the polarization patterns and diffuse scattered 
light in external galaxies (Wood 1997, Wood \& Jones 1997).  These 
investigations adopted smooth axisymmetric models for the dust and 
illuminating starlight.  
In order to determine the propagation of starlight in clumpy media and 
simulate all-sky maps viewed from within a model galaxy, we have modified 
our codes to include the effects of scattering, clumping, dust scaleheights, 
and source distributions in a three dimensional galactic model.  The 
ingredients of our models are the H$\alpha$ emissivity, 
dust density distribution and scattering parameters, and radiation transfer 
algorithm, described below.

\subsection{Density Grid}

Our model galaxy density is constructed on a 3D linear Cartesian 
grid comprising 
$201\times 201\times 201$ grid cells.  The physical extent of our model galaxy 
is $\pm 15$kpc, so each grid cell is a cube of $\approx 150$pc on a side.  
  
The number densities (cm$^{-3}$) for the five phases of the ISM 
(molecular, cold neutral, warm neutral, warm ionized, and hot ionized) are 
determined from analytic expressions presented by 
Ferriere (1998, Eqs.~1 through 9).  
This is a smooth axisymmetric model of the ISM.  In the model at the solar 
radius the 
WIM has a scaleheight of 1kpc and midplane number density of 0.0237cm$^{-3}$.  
The neutral hydrogen is composed of three components: Gaussian scaleheights 
of 127pc and 318pc and an exponential scaleheight of 403pc with the total 
midplane number density (cold neutral plus warm neutral) being 
0.566cm$^{-3}$.  We have also included the 
spiral arm density determined by Taylor \& Cordes (1993).  The spiral 
arms were inserted into our density grid using the FORTRAN subroutines that 
are described in the Appendix of Taylor \& Cordes (1993).
To complete our model ISM density we have inserted an evacuated region of 200pc 
radius centered on the Sun's location.  This approximately represents the 
low density ``local bubble'' (e.g., Cox \& Reynolds 1987), within the 
constraints of the grid size.

We assume that the dust and gas are well mixed in all phases of the ISM 
and that the dust-to-gas ratio is constant throughout the Galaxy.  We have 
not attempted to model any vertical or horizontal dependence of the dust to 
gas ratio within the Galaxy as was speculated by Ferrara et al. (1996).  
This would involve a detailed modeling of many datasets including IRAS, 
H$\alpha$, and H~I and is beyond the scope of our current investigation 
to estimate scattered light levels.  The 
dust scattering parameters are taken to be those of a standard ISM mixture 
(see \S 2.3).  

\subsection{H$\alpha$ Emissivity}

The H$\alpha$ emissivity comprises emission from the WIM 
and that from H~II regions. The WIM H$\alpha$ emissivity is 
proportional to the square of the WIM density (Eq.~7 in Ferriere 1998).  In 
the neighborhood of the Sun and near the Galactic midplane, the mean 
density in the WIM is 0.025~cm$^{-3}$ and the 
emissivity is $6.7\times 10^{-16}$ H$\alpha$ photons/cm$^3$/s (as derived in 
Reynolds 1990).  Extrapolating this emissivity using the Ferriere density 
results in a total WIM emissivity from our model Galaxy of 
$10^{52}$ H$\alpha$ photons/s.  

Bright H~II regions are the dominant source of H$\alpha$ near the Galactic 
equator, with angular sizes ranging from arcminutes to tens of degrees.  
In order to approximate the amount and 
distribution of H$\alpha$ from local H~II regions, we have utilized the 
Garmany et al. (1982) catalog of O stars within 2.5kpc of the Sun.  
This catalog (obtained electronically from the NASA Astrophysical Data Center) 
provides the spectral type of the sources and we convert this to a number of 
ionizing photons from each source using Table~1 of Miller \& Cox (1993).  
We assume 
that 6\% of the ionizing photons from each source escape the Galaxy 
(Bland-Hawthorn \& Maloney 1999) and 15\% go into producing the WIM 
(Reynolds 1990).  Our escape fraction is consistent with the values of 
6\% and 3\% (depending on star formation history) estimated from models of 
Dove et al. 1999.  
The remaining 79\% from each source are assumed to produce local 
``point source'' HII regions with the H$\alpha$ photon luminosity 
equal to $0.46\times 0.79 N_{\rm ionizing}=0.36 N_{\rm ionizing}$, assuming 
Case~B recombination (Martin 1988).  We have assumed that 
$N_{H\alpha}=0.36 N_{\rm ionizing}$ for each source irrespective of 
location.  This is clearly a simplification, since the H$\alpha$ luminosity 
will depend on the local density which in turn determines what fraction of 
ionizing photons escape the Galaxy, what fraction contributes to producing 
the WIM, and what fraction is converted to H$\alpha$ within the local 
H~II region.  This calculation is beyond the scope of our current 
investigation and we therefore globally adhere to the relative fractions 
we have quoted.  The model of Miller \& Cox (1993) attributed the WIM 
H$\alpha$ emission to extended ``Stromgren volumes'' created by OB 
associations and such a scenario would also include H$\alpha$ emission 
from supershells that may have formed around OB associations.  
Note that in our model we have separated the H$\alpha$ into two components: 
a smooth component representing the extended emission of Miller \& Cox and 
the point source H~II regions with H$\alpha$ luminosities as described above.

For regions beyond 2.5kpc (where the Garmany et al. catalog is incomplete) 
we extend the distribution of point sources 
by placing sources randomly in the molecular ring (from Ferriere 1998) and 
in the spiral arms of Taylor \& Cordes (1993).  The (x,y) locations of all 
sources in our simulation are shown in Fig.~1.  We set the H$\alpha$ 
luminosity of each of the additional randomly placed point sources to be 
$10^{49}$ H$\alpha$ photons/s, 
(comparable to that of the Orion Nebula) and add enough sources such that 
the total luminosity of the H~II regions in our simulation is equal to 
that of the WIM, $L_{\rm H~II}=L_{\rm WIM}=10^{52}$ H$\alpha$ photons/s.  
Several authors have estimated the relative contributions 
of H$\alpha$ from the WIM and H~II regions in nearby spiral galaxies, with 
the ratio found to be in the range 
$0.47< {\rm H}\alpha ({\rm H~II})/{\rm H}\alpha ({\rm TOTAL})< 0.7$ for 
each galaxy as a whole (Ferguson, Wyse, \& Gallagher 1996; 
Walterbos \& Braun 1994; Veilleux, Cecile, \& Bland-Hawthorn 1995).  
The relative H$\alpha$ luminosities for the H~II regions and the WIM in our 
simulation have been chosen to match this observational constraint.

\begin{figure}[ht]
\centerline{
\plotfiddle{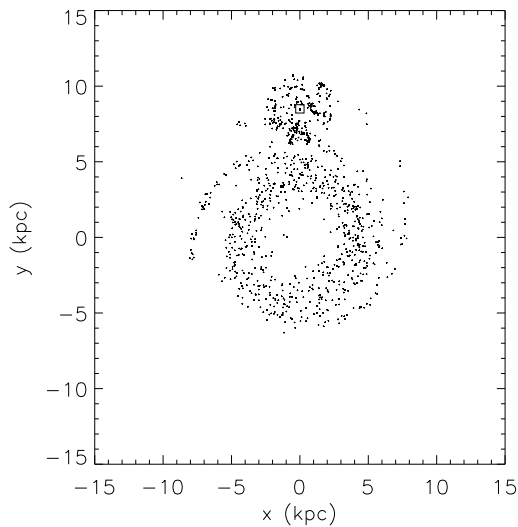}{3.in}{0}{50}{50}{-275}{-150}
}
\caption{The $x$ and $y$ locations of the ``point source H~II regions.'' 
Locations of O stars from the Garmany et al (1982) catalog and random point 
sources distributed in the molecular ring (Ferriere 1998) and in the spiral 
arms of Taylor \& Cordes (1993).  The Sun is located at (0, 8.5, 0).}
\end{figure}

\subsection{Opacity and Scattering Parameters}

We assume that the dust plus gas is represented by a 
Mathis, Rumpl, \& Nordsieck (1977) mixture.  The adopted parameters at 
H$\alpha$ are a total opacity of 
220cm$^2$g$^{-1}$, albedo $a=0.5$, and a scattering phase function 
given by the Heyney-Greenstein phase function, $HG(\theta)$, with $g=0.44$
\begin{equation}
HG(\theta) = {1\over{4\pi}}(1-g^2)/(1+g^2-2g\cos\theta)^{3/2}
\end{equation}
The peak linear polarization, $p_l=0.5$, is taken from White (1979).  
For the adopted densities and opacities, the total optical depth diametrically 
across the galactic equator is $\tau_{\rm eq} \approx 50$, while at the Sun's 
radius, the total optical depth perpendicular to the Galactic disk is 0.4.  

There has been considerable interest in determining dust 
parameters in the diffuse ISM (Murthy \& Henry 1995; Hurwitz 1994; Gordon 
et al. 1994), 
reflection nebulae (Calzetti, et al. 1995), and 
protostellar clouds (Whitney, Kenyon, \& Gomez 1998).  
The effects of changing the albedo and phase function 
asymmetry on scattered light patterns have been presented by other groups.  
Increasing or 
decreasing the albedo obviously increases or decreases the scattered light 
component.  In their model of the UV diffuse Galactic light, Murthy \& Henry 
(1995) showed that increasing $g$ (making the phase function more forward 
throwing) reduces the scattered light levels and concentrates the scattered 
light towards ``halos'' around the UV sources in their simulation.  
Bianchi, Ferrara, \& Giovanardi (1996) 
also pointed out that the Heyney-Greenstein 
phase function may underestimate the forward scattering and this would 
lead to a decrease in scattered light levels.  
Rather than repeat similar calculations for dust scattering of H$\alpha$, 
we note the effects of changing the albedo and phase function and 
keep the dust parameters fixed throughout our simulation.  

\subsection{Radiation Transfer}

In our Monte Carlo radiation transfer we now use a scheme that 
includes 
``forced first scattering'' and a ``peeling off'' procedure as we force the 
photons towards the observer with the appropriate weights (see Appendix).  
We have adopted the 
forced first scattering in order to determine the scattered light contribution 
from scattering in optically thin, high latitude regions.  
Our technique then determines 
the relative contributions of direct and multiple scattered H$\alpha$ 
radiation from the WIM and H~II regions.

One of the major questions in the study of the WIM is the source of its 
ionization.  Current models (Miller \& Cox 1993; Dove \& Shull 1994; 
Dove et al. 1999; Elmegreen 1997) suggest 
that a clumpy ISM may allow photons to leak from the midplane H~II regions 
to the high latitude gas.  While this paper is not investigating this problem, 
we are interested in determining how far the H$\alpha$ probes into the 
Galaxy.  To determine the distance that photons have traveled, we therefore 
construct three all-sky images.  The first contains all photons, 
representing the total H$\alpha$ intensity observed at Earth for this model.  
The two other images contain photons that originated from $>3$kpc and 
$>6$kpc --- using the Monte Carlo technique it is straightforward to 
``label'' photons according to point of origin,  distance traveled, 
number of scatterings, etc.  Thus we 
can determine how far we ``see'' into the galaxy model.  We have also 
separated out direct and scattered H$\alpha$ photons for the WIM and H~II 
regions, thereby enabling us to quantify the role of dust scattering.

\section{Model Results}

Figure~2 shows our all-sky simulation for the ISM model and illumination 
described above.  The bright point sources have been scaled so that they are 
saturated at the maximum intensity from the direct WIM emission in the 
simulation.  In Fig.~3, $18^\circ$ wide latitudinal cuts are shown for 
different Galactic longitudes.  Each panel shows four curves: the heavy line 
is a $\csc |b|$ law and the three other curves show the total H$\alpha$ 
intensity (upper), scattered H$\alpha$ from the point source H~II regions 
(middle), and scattered H$\alpha$ from the WIM (lower).  This 
simulation reproduces many of the observed features of the Galactic H$\alpha$ 
emission.  In particular, the domination of H~II regions at low latitudes, 
the $\csc |b|$ law is closely followed at intermediate latitudes, and the 
intensity is lower than the $\csc |b|$ law at high latitudes.  These gross 
characteristics are all present in the data described by Reynolds (1990).

\begin{figure}[ht]
\centerline{
\plotfiddle{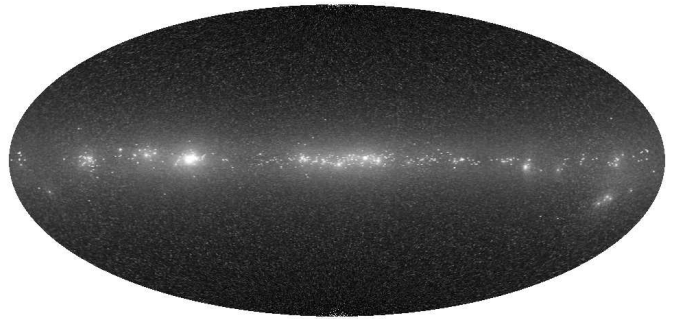}{1.75in}{90}{35}{35}{20}{-20}
}
\caption{All-sky Aitoff projection of our H$\alpha$ intensity simulation 
in Galactic coordinates $(l, b)$.  The Galactic center direction is at the 
center of the figure and Orion is on the right below the midplane.  
The bright point sources have been saturated at the peak level 
of the direct WIM intensity and the WIM and scattered light is shown on a 
square root stretch.}
\end{figure}

\begin{figure}[ht]
\centerline{
\plotfiddle{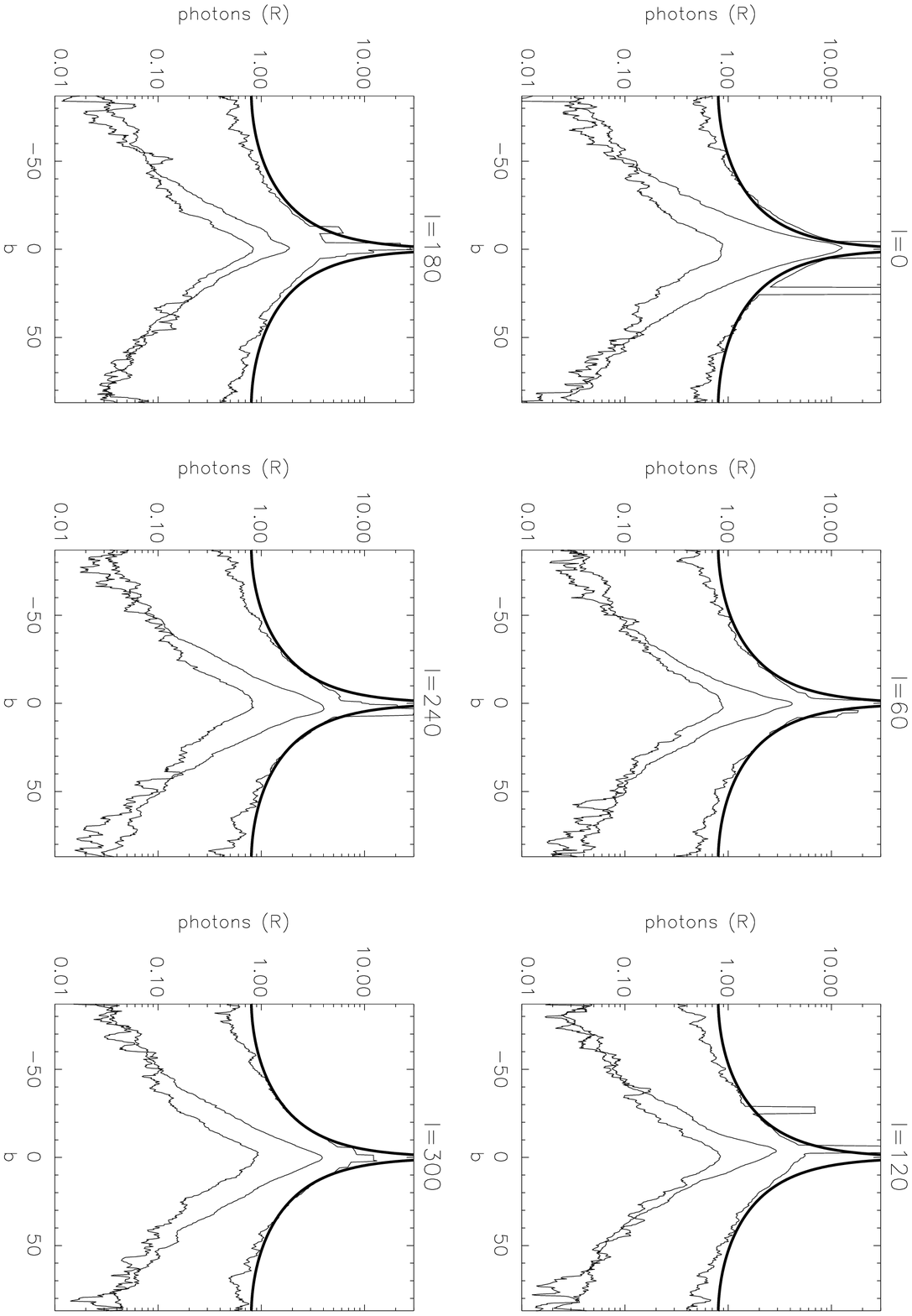}{1.75in}{90}{30}{30}{0}{-30}
}
\caption{$18^\circ$ wide latitudinal intensity cuts through the simulation.  
Solid curve shows a 
$\csc |b|$ law, the three other curves show the total intensity (upper), 
scattered light from point sources (middle), and scattered light from the WIM 
(lower).}
\end{figure}

\begin{figure}[ht]
\centerline{
\plotfiddle{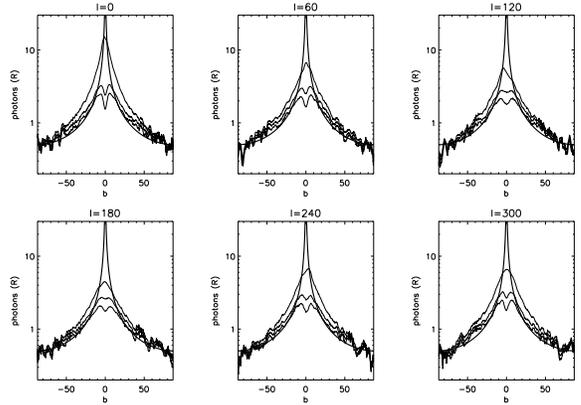}{1.75in}{90}{30}{30}{0}{-30}
}
\caption{Latitudinal cuts through the simulation showing the $\csc |b|$ 
law, and three curves comprising 
different components of the H$\alpha$ intensity.  The upper curve comprises 
the WIM emission (direct plus scattered) and the dust scattered H$\alpha$ from 
the point sources; the middle curve shows the WIM H$\alpha$ (direct plus 
scattered) only; the lower curve shows only the direct WIM emission.}
\end{figure}

\begin{figure}[ht]
\centerline{
\plotfiddle{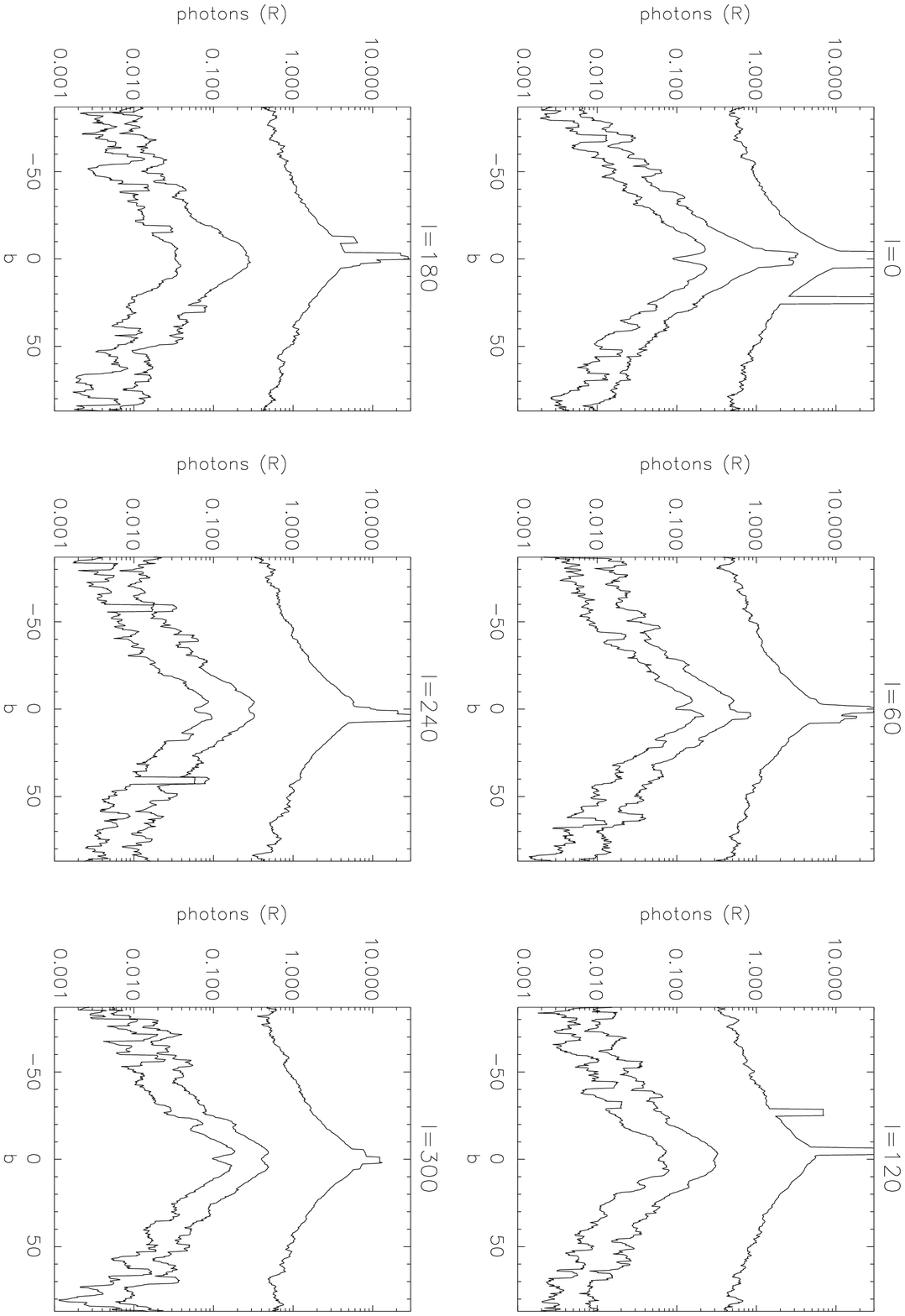}{1.75in}{90}{30}{30}{0}{-30}
}
\caption{Latitudinal cuts through the simulation 
showing the total intensity (upper), H$\alpha$ 
originating from more than 3kpc (middle) and greater than 6kpc (lower).}
\end{figure}

Departures from the $\csc |b|$ law at high latitudes are seen not only in 
the H$\alpha$, but also in H~I observations (Dickey \& Lockman 1990).  
For the H~I, the 
fact that we are located in a local bubble is cited as the explanation for 
the departure.  The local bubble is believed to be elongated perpendicular 
to the Galactic plane (Cox \& Reynolds 1987, Snowden 1986), 
so there is a larger fraction of the H~I column 
density ``carved out'' at high latitudes yielding the apparent departure 
from the $\csc |b|$ law.  The same effect will also be present at H$\alpha$, 
though at a smaller level because of the larger scaleheight of the WIM.  
However, in our simulations we approximated the local bubble as a sphere of 
radius 200pc yet we still found departures from the $\csc |b|$ law at high 
latitudes.  The reason for the departures in our simulations is the larger 
contribution of dust scattered H$\alpha$ at low latitudes.  Figure~4 
illustrates this by presenting a $\csc |b|$ law, and three curves comprising 
different components of the H$\alpha$ intensity.  The upper curve comprises 
the (direct plus scattered) WIM emission and the dust scattered H$\alpha$ from 
the point sources; the middle curve shows the (direct plus scattered) 
WIM H$\alpha$ only; the lower curve shows only the direct WIM emission.  This 
figure clearly shows how scattered H$\alpha$ is larger at low latitudes 
and yields the departures from the $\csc |b|$ law at high latitudes.  Note 
that for the direct WIM emission, the $\csc |b|$ law is closely followed 
at high latitudes, but dust obscuration is responsible for the emission 
being smaller than $\csc |b|$ towards the midplane regions ($|b|<10^\circ$).  
In retrospect the departures are obvious, since the $\csc |b|$ law is 
derived by integrating along paths in a plane parallel layer and does not 
account for any scattered light component.  

An important feature of our simulations is that we may determine the 
scattered light contribution of the H$\alpha$ intensity.  
We find that the scattered light from midplane H~II regions is typically 
around 10\% of the total intensity at high latitudes (Fig.~3).  
Towards the midplane 
the scattered light level is larger due to the proximity of the H~II regions.  
However, our assumptions of a smooth constant density in each grid cell 
likely yields an overestimate of the scattered light for $b<10^\circ$.  In 
reality dust will not survive in the intense radiation environment of the 
O stars and more H$\alpha$ photons will escape these regions without 
scattering than predicted by our model.  Our analysis is primarily concerned 
with estimating the scattered light at high latitudes and a more accurate 
determination of the scattered light at low latitudes requires a three 
dimensional model for the ISM density, in particular in the environs of 
the O stars in our simulation.  Reynolds (1990) estimated the 
scattered light contribution from Galactic H~II regions to be less than 15\% 
based on [S~II]/H$\alpha$ and [O~III]/H$\alpha$ ratios measured along 
sightlines towards H~II regions and along sightlines in the diffuse ISM.  
Our model appears to confirm this estimate at high latitudes.  
The results of our simulation are in 
agreement with the Ferrara et al. (1996) model of NGC~891.  In their 
externally viewed simulation, they estimated an H$\alpha$ scattered light 
contribution of around 10\% at 600pc from the midplane.

We have further separated the H$\alpha$ intensity according to how far 
from the Sun the photons originated.  
Fig.~5 shows panels of latitudinal cuts for the total H$\alpha$ intensity 
(upper), and the intensity of H$\alpha$ photons (direct or scattered) 
originating at distances greater than 3kpc (middle) and 6kpc (lower) from 
the Sun.  It is clear that this simulation is dominated by photons originating 
within 3kpc of the Sun.  
Note the absorption lane in the latitudinal cuts 
of photons from large distances.  This arises because for sightlines above the 
dense dust in the H~I layer the optical depth is low and we see the extended 
diffuse H$\alpha$ from the WIM.  At low latitudes the dust limits how far we 
can see.  It should be possible to see this effect using the kinematic 
information provided by the WHAM survey.  

The effects of dust clumping will modify our 
results that are for a smooth density distribution.  In addition to 
departures from the smooth scattered light cuts we have presented, 
Witt \& Gordon (1996) 
showed that clumping has the overall effect of enabling photons to penetrate 
to larger distances than for the same dust mass arranged in a smooth 
geometry.  The Witt \& Gordon analysis considered radiation transfer 
in a clumpy medium surrounding a point source, which is appropriate for 
the transfer of H$\alpha$ photons from our point sources and will 
likely raise the scattered light levels.  However, since the diffuse WIM 
emission will itself be clumpy we intend to extend the Witt \& Gordon 
analysis to include extended sources of emission appropriate for analysis 
and modeling of the WHAM dataset.  

\section{Polarization Prediction}

With scattered light levels around 10\% (as estimated from line ratio 
measurements and now also from 
our simulations), we would expect this dust scattered H$\alpha$ to 
be polarized.  Therefore, in addition to the scattered light simulations 
shown above, we have also calculated the polarization arising from dust 
scattering.  The H$\alpha$ emission from the WIM and the point sources is 
assumed to be unpolarized and the polarization arises from dust scattering.  
We have assumed that the polarization arises solely from scattering and we 
have not considered the polarizing effect due to the transmission of 
radiation through regions containing aligned dust grains.  Our neglect 
of this transmission polarization is valid for the low column density 
sightlines at high latitude since the transmission polarization scales with 
optical depth (see models by Jones 1989; Jones, Klebe, \& Dickey 1992).  

\begin{figure}[ht]
\centerline{
\plotfiddle{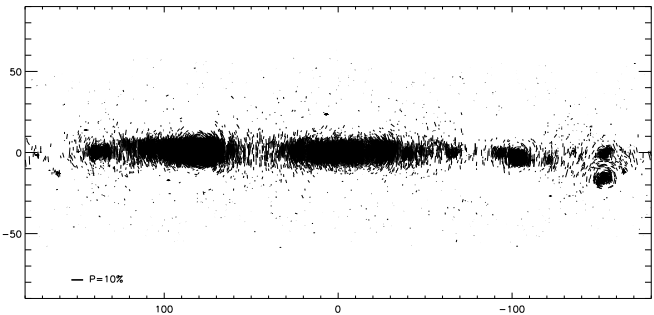}{1.75in}{90}{35}{35}{15}{-30}
}
\caption{Simulated polarization map arising from dust scattering of the 
H$\alpha$ intensity.}
\end{figure}

\begin{figure}[ht]
\centerline{
\plotfiddle{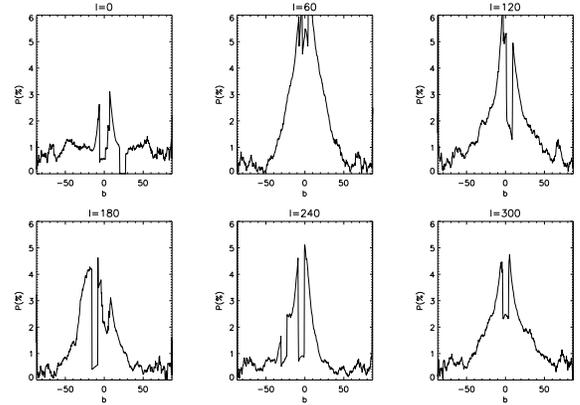}{1.75in}{90}{30}{30}{0}{-30}
}
\caption{$18^\circ$ wide latitudinal polarization cuts.}
\end{figure}

Figure~6 shows an all-sky map showing the predicted polarization 
for our simulation.  The polarization is largest in the midplane regions 
where the amount of scattered light from the point sources is highest.  The 
familiar centrosymmetric patterns are evident indicative of polarization 
arising from scattered light from point sources, or in this simulation 
form concentrations of point sources.  In general the polarization is 
perpendicular to the midplane, as expected for polarization due to scattering.  
Transmission polarization, which is perpendicular to the large scale 
toroidal Galactic magnetic field 
and therefore parallel to the midplane (Mathewson \& Ford 1970, Zweibel 
\& Heiles 1997) will reduce the midplane 
polarization levels.  The combination of low scattered light levels and 
polarization cancelation results in the signal to noise of our simulations 
being too small to allow a detailed prediction of the polarization at high 
latitudes.  To increase the signal to noise, in Fig.~7 
we have taken $18^\circ$ wide latitudinal cuts (as in the previous 
section) and averaged the Stokes parameters over longitude to show the 
average polarization as a function of latitude.  The polarization is at a 
minimum close to the midplane, where the unpolarized direct emission 
dominates.  The polarization then increases at latitudes just above the 
midplane due to the large scattered light levels.  For higher latitudes, where 
our simulations are more accurate, the 
polarization decreases and is predicted to be in the range 0.1\% to 1\%.  
If these predicted polarization levels could be measured they would provide 
another means to probe the distribution of dust, the dust-to-gas ratio, and 
the scattering properties of the dust in the ISM.

\section{Summary}

This paper has presented a simulation of the diffuse H$\alpha$ intensity 
incorporating direct emission from the WIM and the dust scattering of H$\alpha$ 
photons from the WIM and midplane H~II regions.  We have assumed that the 
scattering dust and WIM are distributed according to a smooth 
axisymmetric prescription.  The H~II regions are assumed to be point sources 
coincident with the known locations of O stars within 2.5kpc of the Sun with 
H$\alpha$ luminosities proportional to the ionizing flux from each source.  
For regions beyond 2.5kpc, we have placed point sources randomly into 
spiral arms and the molecular ring, such that the total H$\alpha$ luminosity 
from H~II regions is equal to that from the WIM.

Our simulation appears to reproduce the gross characteristics observed in 
the H$\alpha$ background, namely the $\csc |b|$ law at low latitudes and 
departures from this at high latitudes.  The departures from the $\csc |b|$ 
law are attributed, at least in part, to a larger dust scattered H$\alpha$ 
component from lower latitudes.  We find that dust scattered H$\alpha$ 
from H~II regions contributes less than 20\% of the total intensity, although 
some sightlines towards the midplane dominated by H~II regions show 
higher scattered light levels.  Our models therefore are consistent with 
the line ratio analysis of Reynolds (1990), who estimated the scattered 
light contribution to be less than 15\% of the total intensity.  
We have also made a prediction that the scattered H$\alpha$ should be 
polarized at around the 0.1\% to 1\% level at high latitudes.  

Preliminary results from the WHAM survey show that the H$\alpha$ emission 
is not smooth and there are many filamentary structures present 
(Haffner et al. 1998, Reynolds et al. 1998).  Detailed 
modeling of these data to determine dust to gas ratios, scattering 
phase functions, etc.,  will require a 3D treatment of the ISM density and 
WIM emission, as well as inclusion of large scale Galactic rotation to 
simulate line profiles for the direct and scattered line emission.  
Our codes are set up for this and future modeling efforts 
will be directed towards developing the fully three dimensional models 
demanded by the observations.

\acknowledgements

KW acknowledges support from NASA's Long Term Space Astrophysics Research 
Program (NAG5-6039).  We thank Jon Bjorkman, Barbara Whitney, and Karl Gordon 
for discussions relating to the forced Monte Carlo technique.

\section*{Appendix --- Radiation Transfer Algorithm}

In our previous Monte Carlo investigations we were simulating the transfer 
of photons within an axisymmetric medium and formed images for external 
viewing.  Upon exiting the simulation, scattered light images were formed 
by projecting the photons into $(x,y)$ locations on image planes.  Such 
a technique is efficient for axisymmetric, external viewing situations.  
However, for our current investigation where we are wanting to form images 
viewed from within the density grid, we cannot use this technique since 
photons will rarely scatter in a direction that intersects the 
observer's location.  
We have therefore modified our algorithm as follows 
so that we force weighted photons 
(either direct or scattered) towards the observer, thus enabling us to 
form all-sky images.

H$\alpha$ photons are emitted either from the WIM or from the point source 
H~II regions.  Direct photons are emitted in a direction towards the observer 
and have a weight 
\begin{equation}
W_{\rm direct}={{e^{-\tau}}/{4\pi d^2}}
\end{equation}
where $\tau$ and $d$ are the optical depth and physical distance from 
the point of emission to the observer.

To calculate the scattered light contribution we first choose a random 
direction 
of travel from $4\pi$ steradians and calculate the optical depth, $\tau_1$, 
from the point of emission to the edge of our grid.  We then force the photon 
to scatter at an optical depth less than $\tau_1$, sampled from 
\begin{equation}
\tau=-\log [{1-\xi (1-{\rm e}^{-\tau_1})}]
\end{equation}
this reproduces the correct probability distribution for optical depths and 
ensures that all photons scatter at least once.  
The photons are subsequently scattered into directions that are randomly 
sampled 
from the scattering phase function (Eq.~1).  A new optical depth is then 
generated from $\tau=-\log\xi$ and the scattering location corresponding 
to this optical depth is determined.  If the location is outside our grid, the 
photon is terminated, otherwise a new scattering angle and optical depth are 
generated and the photon is tracked until it exits the grid.  At each 
scattering location we ``peel off'' and direct towards the observer a 
fraction of the photon's energy.  The 
weight of this ``peeled off'' photon is   
\begin{equation}
W_{\rm scatt} = a^N (1-e^{-\tau_1}) {\rm e}^{-\tau_2} HG(\theta) /d_2^2
\end{equation}
where $N$ is the number of scatterings the photon has undergone up to that 
point, $\tau_1$ is the first optical 
depth, and $\tau_2$ and $d_2$ are the optical 
depth and physical distance from the scattering location to the observer.  
The function $HG(\theta)$ weights the photon by the scattering phase function, 
where $\theta$ is the angle through which we force the photon to scatter 
towards the observer.  The total intensity is the sum of the weights of the 
direct (Eq.~2) and scattered (Eq.~3) photons.  We have tested this algorithm 
for external viewing situations against several ``regular'' Monte Carlo codes 
that do not employ any forced scattering or weighting algorithms and find it 
to be accurate and efficient.

\end{document}